\documentstyle[aps,epsf,epsfig,twocolumn,psfig,array]{revtex}

\def\pap{paper}
\begin{document}

\draft \preprint{}

\title{Quantum Statistics Can Suppress Classical Interference}
\author{O. Steuernagel}
\address{Dept. of Physical Sciences,
University of Hertfordshire, College Lane, Hatfield, AL10 9AB, UK}
\date{\today}
\maketitle
\begin{abstract}
Classical optical interference experiments correspond to a measurement of the first-order
correlation function of the electromagnetic field. The converse of this statement: experiments
that measure the first order correlation functions do not distinguish between the quantum and
classical theories of light, does not always hold. A counterexample is given.
\end{abstract} 
\narrowtext
%
\section{Introduction}\label{introduction}
%
Dirac's somewhat unfortunate statement about `interference between two (or more) photons never
occurring` has led to fruitful discussions clarifying the understanding of interference from a
quantum mechanical point of view. These days textbook presentations elucidate that single photon
light fields behave just like classical states of light~\cite{Glauber.coherent} when used in a
Young's double-slit interference experiment~\cite{Walls.buch,Scully.buch,Loudon.buch}. To
exhibit non-classical features, measurements which detect two- or more photon {\em
coincidences}, phase sensitive measurements such as homodyning and quantum
tomography~\cite{Ulf.buch}, and waiting-time distribution (anti-bunching) measurements can be
per\-form\-ed~\cite{Walls.buch,Scully.buch,Loudon.buch,Ulf.buch,Mandel.buch}.

Single photon detection, however, i.e. intensity or first-order coherence
measurements~\cite{1st.Versus.2nd.order}, are often held to be classical in
character~\cite{Walls.buch,Scully.buch}. This conclusion stems from the analysis of Young's
interference patterns for single photon and Glauber-coherent states of light~\cite{Walls.buch},
single photons emitted from two atoms~\cite{Scully.buch} and thermal
fields~\cite{Scully.buch,Loudon.buch}. In these examples quantum and classical predictions for
the form and visibility of interference patterns are
identical~\cite{Walls.buch,Scully.buch,Loudon.buch}. Some physicists have therefore concluded
that "Experiments of this kind which measure the first-order correlation functions of the
electromagnetic field do not distinguish between the quantum and classical theories of
light"~\cite{Walls.buch}. One can even find more generalized statements such as "For fields with
identical spectral properties it is not possible to distinguish the nature of the light source
from only the first-order coherence properties"~\cite{Scully.buch}. Differences between
classical and quantum expressions for first-order coherence phenomena are commonly attributed to
different frequency modes only~\cite{Scully.buch,Loudon.buch}, but we will see that they can
occur for monochromatic fields as well~\cite{monochrom}.

To explain the perceived equivalence between the classical and
quantum behaviour in first-order coherence phenomena some books
refer to Dirac's assertion regarding interference between two
photons not occurring~\cite{Walls.buch,Loudon.buch}. Let us seek a
more detailed explanation instead.

The purpose of this \pap$ $ is twofold. Firstly I will give a simple general proof for the
equivalence between the classical and quantum behaviour in first-order coherence phenomena for
the conventional Young's double-slit setup. And then, I will derive the simplest possible
example based on photon statistics which should show maximal violation of the classically
expected behaviour: whereas in the classical case an interference pattern with perfect
visibility is observed, one should find the same in one but zero visibility in another
corresponding quantum case.
%
\section{Young's double-slit}\label{s2}
%
Firstly, let us briefly reexamine Young's double-slit: it is,
assuming very narrow slits, a pedagogically appealing choice for
explaining classical and quantum interference effects since it is
an intuitive, geometrically simple, pure two-mode system. In its
classic form it has the drawback, though, of not representing the
most general case since both slits are illuminated from the same
source. This limits the choice of states inside the interferometer
to binomially distributed states of light which indeed cannot show
a behaviour different from classical
states~\cite{Glauber.coherent,monochrom}. The proof is given
below.

Giving up on this restriction allows us to tailor the quantum states needed for the
non-classical behaviour of the first-order coherence we want to demonstrate here.

\begin{figure}
\epsfverbosetrue \epsfxsize=2.6in \epsfysize=1.4in
\epsffile[100 600 430 810]{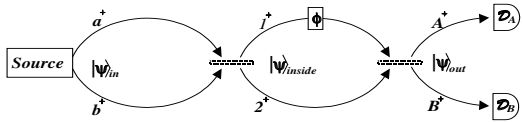} 
\caption{Sketch of the interferometer: dotted lines outline balanced beam splitters, $\cal D$
stands for detectors and $a,b,${\it 1,2},$A$ and $B$ label the modes before, inside and beyond
the interferometer. $\phi$ is the phase shifter in the upper channel. \label{figure1}}
\end{figure}

A Young's double-slit setup can be cast into the general form of a two-mode interferometer
illuminated with two {\em different} fields at its two input ports by being illuminated through
a semitransparent beam-splitter positioned right between its two slits. Such a modified Young's
setup would correspond to a spatially constricted Mach-Zehnder interferometer. In order to avoid
all issues regarding the spatial mode arrangement and parameterization of slit and screen
locations~\cite{monochrom} we will therefore, from now on, consider a general two-mode
(Mach-Zehnder) setup as displayed in FIG.~\ref{figure1}.
%
%
Assuming that both beam-splitters are balanced and equal we choose the modes such that the
action of the first beam splitter is described
by~\cite{Walls.buch,Scully.buch,Loudon.buch,Ulf.buch,Mandel.buch,SalehTeich}
%
\begin{eqnarray}
a^\dagger = ({\mbox{\it 1}^\dagger +  \mbox{\it 2}^\dagger})/{\sqrt 2}, \quad b^\dagger =
({\mbox{\it 1}^\dagger -  \mbox{\it 2}^\dagger})/{\sqrt 2} \; .
 \label{mode.transformation}
\end{eqnarray}
%
The photons following channel {\em 1} are delayed by a tunable phase $\phi$ and are subsequently
mixed with the {\em 2}-channel to form the detector modes $A^\dagger$ and $B^\dagger$:
%
\begin{eqnarray}
\mbox{\it 1}^\dagger = e^{i \phi} \, ({A^\dagger +  B^\dagger})/{\sqrt 2}, \quad \mbox{\it
2}^\dagger = ({A^\dagger -  B^\dagger})/{\sqrt 2} \, ,
 \label{mode.transformation.2}
\end{eqnarray}
%
see Fig.~\ref{figure1}. For the case of a single photon entering the interferometer through mode
$a^\dagger$ we thus receive the final state
%
\begin{eqnarray}
| \psi \rangle_{out} = \frac{(1+e^{i \phi})| 1 \rangle_A | 0 \rangle_B + (-1+ e^{i \phi}) | 0
\rangle_A | 1 \rangle_B }{2}
\label{state.1.out}
\end{eqnarray}
%
and the well known classical photo-detector response probabilities or intensities
%
\begin{eqnarray}
I_A = P_{A}(\phi) & = & \langle A^\dagger  A  \rangle = \frac{1}{2} \; (1+ \cos \phi) \label{signal.1.A} \\
\mbox{  and  } \quad I_B = P_{B}(\phi) & = & \langle B^\dagger B  \rangle = \frac{1}{2} \; (1-
\cos \phi) \; . \label{signal.1.B}
\end{eqnarray}
%
Note that for a single photon entering through mode $b^\dagger$ the role of $I_A$ and $I_B$ or
the signs are exchanged. A simple calculation furthermore confirms that this state is intensity
balanced, that is, inside the interferometer intensities are equal
%
\begin{eqnarray}
I_{\mbox{\it 1}} & = & \langle \mbox{\it 1}^\dagger \mbox{\it 1} \rangle = \frac{1}{2} =
I_{\mbox{\it 2}} = \langle \mbox{\it 2}^\dagger \mbox{\it 2} \rangle\; .
 \label{inside.intensity}
\end{eqnarray}
%
Let us summarize and generalize this result. The input modes 
expressed in terms of
the output modes 
are
%
\begin{eqnarray}
a^\dagger = \frac{(1+e^{i \phi})\; A^\dagger + (-1+ \, e^{i \phi}) \; B^\dagger }{2} \, ,
\label{a.mode.in}
\end{eqnarray}
%
as can easily be read off eq.~(\ref{state.1.out}). The result for $b^\dagger$ is
%
\begin{eqnarray}
b^\dagger = \frac{(-1+e^{i \phi})\; A^\dagger + (1+ \, e^{i \phi}) \; B^\dagger }{2} \, .
\label{b.mode.in}
\end{eqnarray}
%
For a conventional Young's double slit setup the chances of a single photon to pass slit {\it1}
are equal to it passing slit~{\it 2}, for every passing photon $|1\rangle_{into} \mapsto ( |1
\rangle_{\mbox{\it1}} + |1 \rangle_{\mbox{\it2}} )/\sqrt{2} = a^\dagger |0\rangle
$~\cite{Walls.buch,Scully.buch,Loudon.buch}. Every Fock-state component results in a symmetric
binomial distribution of photons, since they randomly pass one hole or the
other~\cite{SalehTeich}. In order to find the general expression for these binomial states, let
us recast this result in the operator language employed in the first equation
of~(\ref{mode.transformation}). We find that an arbitrary pure input state for a conventional
Young's setup has the form
%
\begin{eqnarray}
| \psi \rangle_{in} = \sum_\nu c_\nu \frac{a^{\dagger \nu}}{\sqrt{\nu !}} \, | 0 \rangle \; .
\label{Young.state.in}
\end{eqnarray}
%
Using eq.~(\ref{a.mode.in}) it immediately follows that
%
\begin{eqnarray}
I_A & = & \langle A^\dagger  A  \rangle = \frac{N}{2} \; (1+ \cos \phi)  \; ,
\label{signal.A.young}
\end{eqnarray}
%
where $N$ is the average photon number of $| \psi \rangle_{in}$ in
equation~(\ref{Young.state.in}). Comparing this expression with (\ref{signal.1.A}) we see that
the textbook case of a conventional Young's double-slit indeed always shows classical behaviour.
%
\section{General two-mode interferometers}\label{s3}
%
Let us now study the general case of any pure states occupying modes $a^\dagger$ and $b^\dagger$
simultaneously. In order to get a better understanding for what we are in search of, let us
however briefly remind ourselves of some basics of the theory of optical coherence. The absolute
value of the relative first-order coherence function~$g^{(1)} =
G^{(1)}_{12}/\sqrt{G^{(1)}_{11}G^{(1)}_{22}} $ is connected with the visibility~$V$ of the
interference pattern by~\cite{Walls.buch,Scully.buch,Loudon.buch,Mandel.buch}
%
\begin{eqnarray}
V = |g^{(1)} | \frac{2 \sqrt{I_{\mbox{\it 1}} \, I_{\mbox{\it 2}}}}{I_{\mbox{\it 1}} +
I_{\mbox{\it 2}}} \; . \label{connection.visibility.g1}
\end{eqnarray}
%
In our case of equal intensities this yields the result that visibility equals first order
coherence~\cite{Loudon.buch,Mandel.buch}. With eq.~(\ref{signal.A.young}) we have, in other
words, shown that in a Young's double slit setup 'all' light states~\cite{monochrom} appear to
be first-order coherent, meaning $|g^{(1)}|=1$ everywhere~\cite{Loudon.buch}. Since this is
obviously the largest possible degree of first-order coherence the signatures of a modification
due to quantum statistics can only lead to its {\em suppression}.

In order to study the general case of any pure states occupying modes $a^\dagger$ and
$b^\dagger$ simultaneously we will, for simplicity, expand this state in terms of the modes {\it
1} and {\it 2} {\em inside} the interferometer.
%
\begin{eqnarray}
| \psi \rangle_{inside} = \sum_{\mu \nu} c_{\mu, \nu} \frac{\mbox{\it 1}^{\dagger \mu} \mbox{\it
2}^{\dagger \nu}}{\sqrt{\mu ! \nu !}} \, | 0 \rangle = \sum_{\mu \nu} c_{\mu, \nu} \, | \mu,\nu
\rangle \; .
\label{general.state.in}
\end{eqnarray}
%
With eqs.~(\ref{mode.transformation.2}) we find $A^\dagger A = \frac{1}{2}(\mbox{\it 1}^\dagger
\mbox{\it 1} + \mbox{\it 2}^\dagger \mbox{\it 2} + e^{-i \phi } \mbox{\it 1}^\dagger \mbox{\it
2} + e^{i \phi } \mbox{\it 1} \mbox{\it 2}^\dagger )$ and the corresponding expectation value
for state~(\ref{general.state.in}) is
%
\begin{eqnarray}
\langle A^\dagger A \rangle & = & \sum_{\mu \nu} |c_{\mu, \nu}|^2 \; \frac{\mu+ \nu}{2}
\nonumber
\\
 + & \frac{1}{2} &  \sum_{\mu \nu} \left(  e^{-i \phi } \; c^*_{\mu,\nu} c_{\mu-1,\nu+1}
 \sqrt{\mu ( \nu+1)} \right.
\nonumber \\
 & & \quad \; + \left. e^{i \phi } \; c^*_{\mu,\nu} c_{\mu+1,\nu-1} \sqrt{(\mu +1) \nu } \quad \right)
 \; .
\label{expec.A.general}
\end{eqnarray}
%
Of the state's density matrix only the diagonal terms contribute to the background intensity and
only the first off-diagonal terms determine the interference pattern. To completely erase the
interference pattern we want to get rid of the first off-diagonal terms. For single photon
states that is obviously impossible if one maintains the restriction of balanced illumination of
both interferometric paths. Indeed assuming $|c_{10}|=1/\sqrt{2}$ immediately recovers -- up to
a phase -- our classical result~(\ref{signal.1.A}).

Adding a second photon changes this picture considerably. Let us first consider the conventional
Young's setup again, a two-photon Fock-state incident in mode $a^\dagger$ impinging on the first
beam-splitter $\cal B$ leads to the 'inside' state
%
\begin{eqnarray}
| \psi \rangle_{inside} = {\cal B} \frac{a^{\dagger \, 2}}{\sqrt{2}} | 0 \rangle = \frac{|2,0
\rangle + \sqrt{2} \, |1,1 \rangle + |0,2 \rangle }{2} \; . \label{two.photon.a}
\end{eqnarray}
%
The amplitudes precisely match the corresponding weight factors in the general intensity
expression~(\ref{expec.A.general}) rendering this state first-order coherent. Removal of the $|
1,1 \rangle$-term will completely suppress this first-order coherence. The two-photon state
furthest deviating from the conventional Young's mono-mode input is obviously the balanced
bimodal input state
%
\begin{eqnarray}
| \psi \rangle_{inside} = B  a^{\dagger} b^{\dagger} | 0 \rangle = \frac{|2,0 \rangle - |0,2
\rangle }{\sqrt{2}}  \label{two.photon.a.b}
\end{eqnarray}
%
which, indeed, does not show any first order interference. This state is intensity balanced
inside the interferometer. It therefore only remains to be shown that this state's non-classical
behaviour cannot be attributed to some kind of randomness in phase. Guided by the discussion of
the general intensity expression~(\ref{expec.A.general}) we can expect that a second-order
coherence measurement~\cite{1st.Versus.2nd.order} (two-photon coincidence detection with
probability $P_{AA}$) should give us an interference pattern of higher order since it connects
diagonal density matrix elements with second off-diagonal elements. The corresponding
expectation value for the detection of two photons in channel $A$ has indeed perfect visibility
%
\begin{eqnarray}
P_{AA}(\phi) & = & \frac{1}{2} \langle A^\dagger A^\dagger A A  \rangle =  \frac{1}{4} (1 - \cos
2 \phi)  \; , \label{signal.AA}
\end{eqnarray}
%
here the prefactor $\frac{1}{4}$ normalizes the sum of all detection probabilities per shot to
unity. We note that the effective de Broglie wavelength of this interference effect is halved
with respect to the single photon detection case~\cite{ole.deBroglieWave,Fonseca99}. For our
present considerations the most striking feature is the perfect visibility of this interference
pattern which holds for $P_{AA}$, $P_{BB}$ and $P_{AB}\equiv P_{BA}$ as well. These are the
probabilities to detect two photons in coincidence in the respective channels indicated by the
subscripts~\cite{ole.deBroglieWave}. It is not difficult to show that a second-order pattern
with perfect visibility cannot coexist with a first-order pattern.

With these perfect second-order visibilities state~(\ref{two.photon.a.b}) has to be free of
random phase fluctuations. We are led to consider a state that has a stable relative phase~$
\phi_{\mbox{\it12}}$ across both arms of the interferometer and illuminates both channels of the
interferometer with equal intensity. When we seek a classical description of this state inside
the interferometer before it passes the phase shifter we have to choose a combination of
coherent states~$|\alpha \rangle$ of the form
%
\begin{eqnarray}
|\psi \rangle_{inside} = | \alpha \rangle_{\mbox{\it 1}} |e^{ i \phi_{\mbox{\it \tiny 12} } }
\alpha \rangle_{\mbox{\it 2}} \; . \label{state.should.be.classical}
\end{eqnarray}
%
Classically we would expect this state to be first-order coherent, quantum mechanically,
however, {\em the first-order coherence is suppressed.}

It is clear from our discussion of the general intensity expression~(\ref{expec.A.general})
that, by moving to the second off-diagonal of the density matrix, we have found the simplest
possible case of complete suppression of first-order coherence through quantum statistics.
%
\section{Transition from classical to quantum case}\label{trans}
%
In order to study the transition from classical to quantum behaviour let us look at states that
linearly interpolate ($\eta \in [0,1]$) between the above extremes~(\ref{two.photon.a}) ($\eta =
1$) and~(\ref{two.photon.a.b}) ($\eta = 0$)
%
\begin{eqnarray}
| \psi \rangle_{in} = \left( \sqrt{\eta} \, \frac{ a^{\dagger 2} }{\sqrt{2}}  + \sqrt{1-\eta} \;
a^\dagger b^\dagger \right) | 0 \rangle \; . \label{interpolation.state.in}
\end{eqnarray}
%
These states' first-order visibilities are $V_Q=\eta$, as can be seen from
%
\begin{eqnarray}
I_{A} & = &  1 +\eta \cos \phi  \; 
\mbox{  and  } \; I_{B} = 2 - I_{A} \; . \label{signal.compare.Q.C.B}
\end{eqnarray}
%
The respective channel intensities inside the interferometer are no longer balanced
%
\begin{eqnarray}
I_{\mbox{\it 1}} & = &  1-\sqrt{ 2 \eta (1-\eta)} \; 
\mbox{  and  } \; I_{\mbox{\it 2}} = 2 - I_{\mbox{\it 1}} \; ; \label{signal.compare.Q.C.2}
\end{eqnarray}
%
we will therefore have to compare classical and quantum behaviour using a channel
intensity-independent measure, namely the ratio of first order coherence functions
$|g_Q^{(1)}|/|g_C^{(1)}|$, where the subscripts stand for 'quantum' and 'classical' case.
Because of the stable relative phase~$ \phi_{\mbox{\it12}}$ across both arms of the
interferometer and our assumed perfect mode match~\cite{monochrom} we expect
$|g_C^{(1)}|=1$~\cite{Loudon.buch,Mandel.buch}. From eq.~(\ref{connection.visibility.g1}) we can
hence infer that the expected classical visibility only depends on the channel intensities, $V_C
= {2 \sqrt{I_{\mbox{\it 1}} \, I_{\mbox{\it 2}}}}/(I_{\mbox{\it 1}} + I_{\mbox{\it 2}}) $, which
are given above. Dividing the visibilities yields $V_Q/V_C = |g_Q^{(1)}|/|g_C^{(1)}| =
|g_Q^{(1)}|$. The modulus of the quantum-statistically suppressed first-order coherence function
%
\begin{eqnarray}
|g_Q^{(1)} (\eta) | =
 \frac{\eta}{
 \sqrt{(1-\sqrt{2\eta\, (1-\eta)}) \, (1+\sqrt{2\eta\, (1-\eta)}) } } \; ,
 \label{quantum.g.1}
\end{eqnarray}
%
falls far below the classically expected value of '1' when we approach the bimodal input
state~(\ref{two.photon.a.b}) at $\eta =0$.
%
\begin{figure}
\epsfverbosetrue \epsfxsize=3.0in \epsfysize=1.4in
%
\epsffile{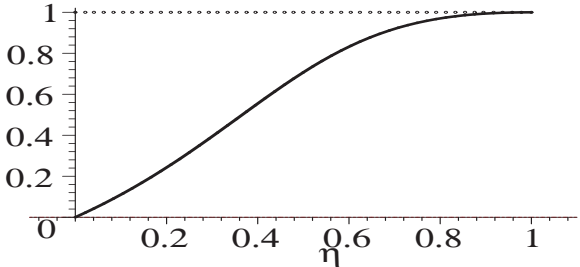}
\caption{Plot of the modulus of the quantal first-order coherence function $|g_Q^{(1)} (\eta) |$
of state~\protect{(\ref{quantum.g.1})} compared with the classical value $|g_C^{(1)} (\eta)
|=1$, dotted line.
%
\label{figure2}}
\end{figure}
%
In the quantum-classical comparison leading to FIG.~\ref{figure2} we assumed the same channel
intensities $I_{\mbox{\it 1}}$ and $I_{\mbox{\it 2}}$ and stability of the
cross-phase~$\phi_{\mbox{\it 12}}$. One can compare the classical to the quantum case employing
a different set of assumptions. We could for example keep the input intensities $I_a$ and $I_b$
matched and also vary the relative phase between modes $a^\dagger$ and $b^\dagger$ by, say,
multiplying the latter with a phase factor: $e^{i \beta} b^\dagger$. Input state $|\psi_{in}
\rangle$ of eq.~(\ref{interpolation.state.in}) now depends on $\eta $ and $\beta$ and so does
its classical counterpart
%
\begin{eqnarray}
|\psi \rangle_{inside} = \left| \sqrt{1+\eta} \right\rangle_{\mbox{\it 1}} \left|e^{ i \beta }
\sqrt{1-\eta} \right\rangle_{\mbox{\it 2}} \; . \label{interpolation.state.classical}
\end{eqnarray}
%
Here the Glauber-coherent state amplitudes~\cite{Glauber.coherent} match the input intensities
$I_a=1+\eta$ and $I_b=2-I_a$ of state~(\ref{interpolation.state.in}). The interference signals
seen by detector $A$ for the phase tunable version of~(\ref{interpolation.state.in}) are
%
\begin{eqnarray}
I_{A,Q} (\eta,\beta) = 1 + \eta \cos \phi - \sqrt{2 \eta \, (1-\eta)} \sin \beta \sin \phi
 \; \label{I.A.Q}
\end{eqnarray}
%
and
%
\begin{eqnarray}
I_{A,C} (\eta,\beta) = 1 + \eta \cos \phi - \sqrt{(1+\eta)(1-\eta)} \sin \beta \sin \phi \; .
\label{I.A.C}
\end{eqnarray}
%
for its classical counterpart~(\ref{interpolation.state.classical}). The resulting visibilities
are the pythagorean combinations of the trigonometric coefficients
%
\begin{eqnarray}
V_{Q} (\eta,\beta) & = & \sqrt{ \eta^2 + 2 \eta (1-\eta) \sin^2 \beta }
\label{V.A.Q} \\
\mbox{and} \quad V_{C} (\eta,\beta) & = &  \sqrt{ \eta^2 + (1+\eta)(1-\eta) \sin^2 \beta }\; .
\label{V.A.C}
\end{eqnarray}
%
The quantum-statis\-ti\-cal sup\-pression $V_{Q} (\eta,\beta) \leq$ $ V_{C} (\eta,\beta)$ is
again strongest for $\eta =0$.

The two classical models, just presented, are incompatible with each other since the second
model's input intensities $I_a$ and $I_b$ do not yield the first model's channel intensities
$I_{\mbox{\it 1}}$ and $I_{\mbox{\it 2}}$ and vice versa. Consequently this discussion is to
some extent arbitrary but it still serves to show that a physically reasonable classical model
cannot explain the quantum-statistical suppression of first-order coherence.
%
\section{Preparing the state} \label{preparing}
%
For experimental confirmation of the results presented here it is important to realize how easy
it is to prepare the interpolation state, let us therefore rewrite
eq.~(\ref{interpolation.state.in}) with $\eta$ as a mixing angle
%
\begin{eqnarray}
| \psi \rangle_{in} = \left( \cos {\eta} \, \frac{ a^{\dagger 2} }{\sqrt{2}}  + \sin {\eta} \;
a^\dagger b^\dagger \right) | 0 \rangle \; . \label{interpolation.state.in.angle}
\end{eqnarray}
%
This form can be synthesized using spontaneous parametric
down-conversion~\cite{Walls.buch,Scully.buch,Mandel.buch} and passive linear devices. Let us
think of a configuration where signal~$\mbox{\it 1}^\dagger$ and idler~$\mbox{\it 2}^\dagger$
photons of a single pair are mode-matched except for orthogonal polarization, say horizontal
'$H$' and vertical '$V$'. The initial state is thus in the state $\mbox{\it 1}^\dagger_H
\mbox{\it 2}^\dagger_V | 0 \rangle$. Next the $\mbox{\it 2}$-mode's polarization is turned such
that it overlaps $\mbox{\it 1}$ by an amount $\cos \eta$, subsequently the two spatial modes
$\mbox{\it 1}$ and $\mbox{\it 2}$ are mixed with a polarization-insensitive beam splitter
according to transformation~(\ref{mode.transformation.2}). This yields the combination
%
\begin{eqnarray}
| \psi \rangle & = & \frac{A_H^\dagger +B_H^\dagger }{\sqrt 2} \cdot \frac{
\cos \eta (A_H^\dagger
-B_H^\dagger ) + \sin \eta (A_V^\dagger -B_V^\dagger )
}{\sqrt 2} | 0 \rangle
\\ & = &
%
\left[ \frac{a^\dagger }{\sqrt 2}  \left(  \cos \eta \frac{a^\dagger }{\sqrt 2} + \sin \eta
b^\dagger  \right) \right.
\left. - \frac{c^\dagger }{\sqrt 2}  \left( \cos \eta \frac{c^\dagger }{\sqrt 2} - \sin \eta
b^\dagger \right) \right] | 0 \rangle
\; , \label{prep.state.in.angle}
\end{eqnarray}
%
where the identifications $A_H \mapsto a$, $(A_V-B_V)/\sqrt{2} \mapsto b $ and $B_H \mapsto c $
have been made. The presence of the $c^\dagger $ photon allows us to identify the unwanted
$c^\dagger b^\dagger $-component which otherwise would pollute the contribution from the desired
first two terms which constitute state~(\ref{interpolation.state.in.angle}).
%
\section{Conclusion}\label{conclusion}
%
To conclude, we have given a general proof for the equivalence of the first-order coherence
function for the classical and the quantum case measured with a conventional Young's double-slit
setup. We have shown that this equivalence does not hold for arbitrary two-mode interferometers
and have derived the simplest possible case for maximal quantum-statistical suppression of
first-order coherence. This case was analyzed by comparing it to two classical scenarios.
Finally have we shown how to synthesize the quantum states used in this paper in order to allow
for experimental confirmation of the predicted quantum-statistical suppression of first-order
coherence.

\acknowledgments I wish to thank Janne Ruostekoski for discussions.


\begin{thebibliography}{99}

\bibitem{Glauber.coherent} By 'classical' I mean 'behaving like a Glauber-coherent state'.

\bibitem{Walls.buch} D. F. Walls and G. J. Milburn, {\em Quantum Optics},
(Springer, Berlin, 1995).

\bibitem{Scully.buch} M. O. Scully and M. S. Zubairy, {\em Quantum Optics}, (Cambridge Univ.
Press, Cambridge, 1997).

\bibitem{Loudon.buch} R. Loudon, {\em The Quantum Theory of Light},
(Clarendon Press, Oxford, 1984).

\bibitem{Ulf.buch} U. Leonhardt, {Measuring the Quantum State of Light}, (Cambridge Univ.
Press, Cambridge, 1997).

\bibitem{Mandel.buch} L. Mandel and E. Wolf, {\em Optical Coherence and Quantum Optics},
(Cambridge Univ. Press, Cambridge, 1995).

\bibitem{1st.Versus.2nd.order} I use the terminology of
references~\cite{Walls.buch,Scully.buch,Loudon.buch} and refer to $g^{(1)}$ as a {\em first}
order function since it is of first order in the measured intensities, this is at variance with
the terminology in other parts of the literature~\cite{Mandel.buch} where it is refered to as a
function of {\em second} order, in terms of amplitudes.

\bibitem{monochrom} I will only consider pure states populating single spatio-temporal
modes~\cite{Titulaer66} (quasi-monochromatic, one polarization) of sufficient transversal
coherence length as to coherently illuminate both slits. In the case of paths overlapping at the
beam splitters I always assume perfect matching of these modes. The generalization to mixtures
of such pure states is straightforward~\cite{Loudon.buch}.

\bibitem{Titulaer66} U. M. Titulaer and R. J. Glauber,
Phys. Rev. {\bf 145}, 1041 (1966).

\bibitem{SalehTeich} R. A. Campos, B. E. A. Saleh, and M. C. Teich,
Phys. Rev. A {\bf 40}, 1371 (1989).

\bibitem{ole.deBroglieWave} O. Steuernagel, {\em De Broglie Wavelength Reduction
for a Multi-photon Wave Packet}, submitted.

\bibitem{Fonseca99} E. J. S. Fonseca, C. H. Monken, and S. Padua,
Phys. Rev. Lett. {\bf 82}, 2868 (1999).

\end{thebibliography}
\end{document}